\documentclass[5p,times,twocolumn]{elsarticle}

\usepackage[english]{babel}
\usepackage[utf8]{inputenc}
\usepackage{amsmath}

\usepackage{color}
\biboptions{sort&compress}

\DeclareMathOperator{\Tr}{Tr}

\begin{document}

\begin{frontmatter}
\title{Conserved Charge Fluctuations in a Chiral Hadronic Model
  including Hadrons and Quarks}

\author[1,2]{P.\ Rau}%
\ead{rau@th.physik.uni-frankfurt.de}%
\author[3]{J.\ Steinheimer}%
\author[1,2]{S.\ Schramm}%
\author[1,2,4]{H.\ St\"ocker}%
\address[1]{Institut f\"ur Theoretische Physik, Goethe Universit\"at,
  Max-von-Laue-Str.\ 1, 60438 Frankfurt am Main, Germany}%
\address[2]{Frankfurt Institute for Advanced Studies (FIAS),
  Ruth-Moufang-Str.\ 1, 60438 Frankfurt am Main, Germany}%
\address[3]{Lawrence Berkeley National Laboratory, Berkeley, CA 94720,
  USA}%
\address[4]{GSI Helmholtzzentrum f\"ur Schwerionenforschung GmbH,
  Planckstr.\ 1, 64291 Darmstadt, Germany}%


\begin{abstract}
  In this work the baryon number and strange susceptibility of second
  and fourth order are presented. The results at zero baryonchemical
  potential are obtained using a well tested chiral effective model
  including all known hadron degrees of freedom and additionally
  implementing quarks and gluons in a PNJL-like approach. Quark and
  baryon number susceptibilities are sensitive to the fundamental
  degrees of freedom in the model and signal the shift from massive
  hadrons to light quarks at the deconfinement transition by a sharp
  rise at the critical temperature. Furthermore, all susceptibilities
  are found to be largely suppressed by repulsive vector field
  interactions of the particles. In the hadronic sector vector
  repulsion of baryon resonances restrains fluctuations to a large
  amount and in the quark sector above $T_c$ even small vector field
  interactions of quarks quench all fluctuations unreasonably
  strong. For this reason, vector field interactions for quarks have
  to vanish in the deconfinement limit.
\end{abstract}

\begin{keyword}
  chiral effective model \sep QCD phase transition \sep conserved
  charge fluctuations \sep susceptibilities%
  \PACS 25.75.-q \sep 0.Rd \sep 25.75.Nq \sep 24.60.Ky%
\end{keyword}

\end{frontmatter}
\section{Introduction}
\label{sec:introduction}

A major objective of heavy-ion experiments as performed at the RHIC,
LHC, and future experiments at the upcoming Facility for Antiproton
and Ion Research (FAIR) is to study properties of strongly interacting
matter, particularly characteristics of the phase transition at high
temperatures and baryon densities. There are robust indications that
in high-energy nuclear collisions an extremely hot and dense state of
matter forms. This quark–gluon plasma (QGP) shows characteristics of a
nearly perfect fluid with very low
viscosity~\cite{arsene_quarkgluon_2005, back_phobos_2005,
  adams_experimental_2005, adcox_formation_2005}. Lattice QCD has
found this deconfinement transition from a hadron resonance gas (HRG)
to a gas of quarks and gluons at zero baryonchemical potential $\mu_B
= 0$ to happen at $T \approx 160$~MeV in a smooth cross-over for all
thermodynamic variables~\cite{aoki_order_2006,
  bazavov_chiral_2011}. At finite $\mu_B$, this phase transition is
shifted to smaller temperatures~\cite{de_forcrand_chiral_2008,
  endrodi_qcd_2011, bazavov_chiral_2011, kaczmarek_phase_2011},
whereas the exact position, the order of the phase transition, and the
potential existence of a critical endpoint in the region $\mu_B > 0$
are still subject of scientific study.\par
The extraction of robust observables for the phase transition from
final state particles remains a major difficulty in studying the QCD
phase diagram experimentally. Since average fluctuations of quantum
numbers in a finite volume differ significantly between the confined
and deconfined phase, fluctuations of conserved charges, such as of
the net baryon number and the electric charge, are suitable indicators
for the phase transition and may signal QGP
formation~\cite{asakawa_fluctuation_2000, jeon_charged_2000}.\par
Generally, in experiments fluctuations of observables occur due to
systematic uncertainties in experimental techniques including inexact
measurement processes and statistical uncertainties. Additional random
fluctuations of more fundamental nature exist which can be attributed
to the dynamics and thermodynamics of the system under
consideration. In the microscopic limit, random density fluctuations
occur in an early stage of a dynamically evolving system. In heavy-ion
collisions initial event-by-event inhomogeneities arise due to
randomly distributed impact parameters and colliding nucleons as well
as to quantum fluctuations in the scattering cross
sections~\cite{stephanov_event-by-event_1999}. Initial inhomogeneities
in a thermalized system can strongly be amplified, when the system
evolves through a phase transition~\cite{chomaz_nuclear_2004,
  steinheimer_spinodal_2012, steinheimer_spinodal_2013}. At the
critical temperature of a first-order phase transition, two degenerate
thermal equilibriums exist. When the hot system cools down through the
transition, parts of the matter can remain in an unstable local
minimum and due to spinodal decomposition narrowly defined regions
with different thermodynamic properties can emerge. Furthermore,
processes such as critical slowing down, reheating of the system, and
the formation of domains can occur in a dynamical system near a
critical point or when crossing a phase
transition~\cite{stephanov_signatures_1998,
  stephanov_event-by-event_1999, nahrgang_nonequilibrium_2011,
  herold_chiral_2013, bleicher_fluid_2013}.\par
Amplifications of fluctuations at the phase transition are likely to
occur in heavy-ion collision, in which a highly excited and heated
fireball (with parts of it potentially in thermal equilibrium) expands
and cools down.  In this dynamic process matter may cross a first or
higher-order phase transition and potential clumping of matter and the
impact of density fluctuations during hadronization may be
recognizable in particle observables such as a possibly higher
production rate of heavy fragments and of exotic nuclei. Therefore,
the enhancement of (event-by-event) fluctuations may hint the creation
of a QGP~\cite{asakawa_fluctuation_2000, jeon_charged_2000}, hitting
the critical point~\cite{stephanov_signatures_1998,
  stephanov_event-by-event_1999, hatta_proton-number_2003,
  hatta_universality_2003, stephanov_qcd_2005} or crossing a
first-order phase transition~\cite{nahrgang_nonequilibrium_2011,
  herold_chiral_2013, bleicher_fluid_2013, steinheimer_spinodal_2013,
  steinheimer_spinodal_2012}. However, no experimental indications for
increased fluctuations in the transverse momentum close to the
suggested region of a critical end-point have been observed
yet~\cite{adams_event-wise_2005, grebieszkow_event-by-event_2007,
  adamova_scale-dependence_2008}.\par
In heavy-ion experiments, fluctuations can be best studied on an
event-by-event basis. Observations of fluctuations are restricted to
accessible observables, such as correlations in the momenta of
produced particles and fluctuations of the quantum numbers in small
sub-volumes. In theoretical models assuming thermal equilibrium,
fluctuations of conserved charges in defined volumes, such as electric
charge, baryon number, (strange) quark number, and other quantum
numbers, are known to correlate with higher-order cumulants of the
partition function, so-called susceptibilities. This approach to
fluctuations is widely used in particular by lattice QCD and other
theoretical models for strongly interacting
matter~\cite{schmidt_conserved_2013, skokov_volume_2012,
  redlich_probing_2012,bazavov_fluctuations_2012,
  friman_fluctuations_2011}.\par
Relating theory considerations with experiment,
in~\cite{koch_hadronic_2008} it is pointed out that susceptibilities
can be measured experimentally ``since they can be expressed as
integrals over either spatial or momentum space correlation
functions. Thus, as long as one deals with susceptibilities, i.e
(co)-variances, there is a one to one mapping from lattice QCD results
to heavy-ion collisions [$\ldots$].
The susceptibilities can be extracted from data either by studying
event-by-even fluctuations of a given quantity or by measuring and
integrating the appropriate multi-particle
densities~\cite{bialas_event-by-event_1999}.'' This implies the
comparability of experimental data to theoretical models, such as PNJL
and chiral effective models.\par
For the outlined reasons, susceptibilities from the chiral model
should signal the shift in the degrees of freedom at the phase
transition due to the drop in the effective baryon masses (chiral
transition) as well as the rising quark abundance above $T_c$
(deconfinement transition). Comparing model results to lattice QCD can
give insight into potential differences in the underlying degrees of
freedom and to what extent this transition is driven by hadrons or
quarks.

\section{Chiral Effective Model}
\label{sec:chiral-effective-model}

This work studies fluctuations of conserved charges at the phase
transition using a unified approach to QCD matter. The effective model
combines a SU(3)-flavor $\sigma$-$\omega$
model~\cite{boguta_systematics_1983, papazoglou_chiral_1998,
  papazoglou_nuclei_1999, dexheimer_proto-neutron_2008} with a
PNJL-type approach for deconfinement~\cite{fukushima_chiral_2003,
  meisinger_coupling_1995, meisinger_chiral_1995, ratti_phases_2006,
  ratti_phase_2006, ratti_thermodynamics_2006,
  roessner_polyakov_2006}. The model features a chiral and
deconfinement transition and includes both a HRG phase with the
spectrum of all known hadrons with masses $m_H \le
2.6$~GeV~\cite{particle_data_group_review_2012, nakamura_review_2010}
as well as a quark-gluon phase at high temperatures and densities. In
the following, basic concepts of the model are shortly outlined;
see~\cite{rau_chiral_2013} for a comprehensive review and all
parameter values.\par
In mean field approximation~\cite{serot_relativistic_1986,
  serot_recent_1997} the full Lagrangian reads $\mathcal{L} =
\mathcal{L}_{\rm kin} + \mathcal{L}_{\rm int} + \mathcal{L}_{\rm
  mes}$. It includes the kinetic energy of the hadrons
$\mathcal{L}_{\rm kin}$~\cite{papazoglou_nuclei_1999}. Furthermore,
$\mathcal{L}_{\rm int}$ describes the attractive interaction of
baryons and quarks with the scalar isoscalar mesons condensates
$\sigma$, $\zeta$ and the repulsive interaction with the vector
isoscalar fields $\omega$, $\phi$ expressed by
\begin{equation}
  \label{eq:L_int}
  \mathcal{L}_{int} = -\sum_i \bar{\psi_i} \left[  \gamma_0 \left( g_{i\omega}
      \omega^0 + g_{i\phi}
      \phi^0 \right) + m^*_i \right] \psi_i .
\end{equation}
Index $i$ runs over the three lightest quark flavors ($u$, $d$, $s$),
the baryon octet, decuplet, and all heavier baryon resonances. The
$\sigma$-field is the order parameter for the chiral transition.\par
Except for a small explicit mass $\delta m_i$, the particles' coupling
strengths $g_{i \sigma, \zeta}$ to the scalar fields dynamically
generate the effective masses
\begin{equation}
  \label{eq:effective_mass}
  m_{i}^* = g_{i\sigma}\sigma + g_{i\zeta}\zeta + \delta m_i .
\end{equation}
It is $\delta m_{u, d} = 6$~MeV, $\delta m_s = 105$~MeV for the quarks
and $\delta m_i = 150$~MeV for nucleons. The value of $\delta m_i$
becomes larger with increasing vacuum mass of the specific particle. A
decreasing $\sigma$-field at high $T$ and $\mu$ causes the effective
baryon masses to drop and, thus, chiral symmetry to be
restored. Accordingly, the effective chemical potentials for quarks
and baryons $\mu^*_i = \mu_i - g_{i \omega} \omega - g_{i \phi} \phi$,
are generated by the vector couplings $g_{i \omega, \phi}$.\par
Couplings of the baryon octet are fixed such as to reproduce
well-known vacuum masses, nuclear saturation properties, and the
asymmetry
energy~\cite{dexheimer_proto-neutron_2008,dexheimer_novel_2009},
resulting in $g_\sigma^N = -9.83$, $g_\zeta^N = -1.22$, $g_\omega^N =
11.56$ for the nucleons. Quark couplings $g_\sigma^{u,d} = -3.5$ for
non-strange quarks and $g_\zeta^s = -3.5$ for strange quarks are fixed
such as to restrain free quarks from the ground state and to comply
with the additive quark model. The quark vector couplings $g_{qv}$,
i.e.\ $g_\omega^{u,d}$ and $g_\phi^s$, remain free parameters to study
the repulsive effect of vector field interactions.\par
The baryon resonance couplings (including the decuplet) are scaled by
$r_s$, $r_v$ to the respective couplings of the nucleons via $g_{B_i
  \sigma, \zeta } = r_{s} \cdot g_{N \sigma, \zeta}$ and $g_{B_i
  \omega, \phi} = r_{v} \cdot g_{N \omega,
  \phi}$~\cite{rau_baryon_2012}.  To obtain a cross-over at $\mu_B =
0$ in all quantities, the scalar resonance coupling is fixed $r_s
\approx 1$. The vector coupling $r_v$ is varied in order to study the
suppressive effect of vector field interactions. Baryon resonances
have large impact on the overall phase structure and the resulting
order and position of the phase transition. Reasonably large resonance
vector couplings rule out a potential first-order phase transition in
favor for a smooth cross-over in the whole $T$-$\mu$
plane~\cite{rau_baryon_2012}. This smooth transition is due to the
gradual population of heavy-mass resonances states. A non-interacting
HRG is considered by neglecting all particle interactions with the
fields and setting $\delta m_i$ to the respective vacuum mass.\par
The meson part of the full model Lagrangian {\allowdisplaybreaks
\begin{align}
  \begin{split}
    \label{eq:self_int_vec_mes}
    \mathcal{L}_{\rm mes} &= \frac{1}{2} \frac{\chi}{\chi_0} \left(
      m^2_{\omega} \omega^2 + m^2_{\phi} \phi^2 \right)\\
    & \quad +g_4 \left( \omega^4 + \frac{\phi^4}{4} + 3 \omega^2
      \phi^2 + \frac{4 \omega^3 \phi}{\sqrt{2}} + \frac{2 \omega
        \phi^3}{\sqrt{2}} \right)\\
    & -\frac{1}{2} k_0' \, \left( \sigma^2 +
      \zeta^2 \right) + k_1 \, \left( \sigma^2 + \zeta^2 \right)^2 \\
    &\quad + k_2 \, \left( \frac{\sigma^4}{2} + \zeta^4 \right) + k_3'
    \,\sigma^2 \zeta\\
    & \quad- k_4 \, \chi^4 - \frac{1}{4} \chi^4\, \ln
    \frac{\chi^4}{\chi_0^4} + \frac{\delta}{3}\, \chi^4 \ln{
      \frac{\sigma^2 \zeta} {\sigma_0^2 \zeta_0} }\\
    & - \frac{\chi^2}{\chi_0^2} \left[ m_\pi^2
      f_\pi\sigma+\left(\sqrt{2}m_k^ 2f_k -\frac{1}{\sqrt{2}}m_\pi^ 2
        f_\pi\right)\zeta \right]
  \end{split}
\end{align}}%
includes the mass terms, self interactions of the vector and scalar
mesons, and explicit symmetry breaking. In the absence of quarks, the
dilaton field $\chi$, introduced as a gluon condensate in order to
ensure QCD scale invariance~\cite{papazoglou_nuclei_1999}, is fixed at
its ground state value $\chi_0$. In order to suppress the chiral
condensate in the deconfined quark phase, a coupling of the Polyakov
loop $\Phi$ to the dilaton field is introduced via
\begin{equation}
  \label{eq:coupling-polchi}
  \chi = \chi_0\left[ 1 - 1/4\, \left( \Phi^2 + \bar{\Phi}^2 \right)^2
  \right].
\end{equation}
All thermodynamic quantities are derived from the grand canonical
potential
\begin{equation}
  \label{eq:grand_canon_pot}
  \Omega / V = -\mathcal{L}_{\rm int} - \mathcal{L}_{\rm mes}
  + \Omega_{\rm th} / V  - U_{\rm Pol},
\end{equation}
with $\Omega_{\rm th}$ defined in the heat bath of hadrons and quarks
including thermal contributions from mesons, baryons, and quarks
\begin{align}
  \label{eq:gc_pot_thermal_qq}
  \Omega_{\rm q\bar{q}} = &- T \sum_j \frac{\gamma_j}{(2
    \pi)^3} \int d^3k \; \left( \ln \left[ 1 + \Phi\, e^{ -\frac{1}{T}
        \left( E^*_j(k) - \mu^*_j
        \right)}\right] \right.\nonumber \\
  &\left. +\ln \left[ 1 + \bar{\Phi}\, e^{ -\frac{1}{T} \left(
          E^*_j(k) + \mu^*_j \right)} \right] \right),
\end{align}
with $j = u,d,s$, the spin-isospin degeneracy factor $\gamma_j$, and
the single particle energy $E^*_j(k) = \sqrt{k^2 + m_j^{*2}}$.\par
Quarks are introduced in the style of recent PNJL
models~\cite{fukushima_chiral_2003, meisinger_coupling_1995,
  meisinger_chiral_1995, ratti_phases_2006, ratti_phase_2006,
  ratti_thermodynamics_2006, roessner_polyakov_2006} defining the
scalar Polyakov loop field $\Phi$ via the trace of the time component
$A_0$ of the SU(3) color gauge background field $\Phi = 1/3\,
\Tr{\left[ \exp{\left( - A_0 / T \right) } \right] }$.  In the
heavy-quark limit, $\Phi$ signals the breakdown of Z(3) center
symmetry and serves as an order parameter for deconfinement. The
transition dynamics from the HRG to the deconfined quark-gluon phase
are controlled by the effective Polyakov loop potential
\begin{align}
  \label{eq:Polyakov-loop-eff-pot}
  \begin{split}
    U = - \left( a(T) \bar{\Phi} \Phi \right) / 2 + b \left( T_0 / T
    \right)^3 \ln \left[ 1 - 6\bar{\Phi}\Phi \right.\\
    \left. + 4 ( \bar{\Phi}^3 + \Phi^3 ) - 3 ( \bar{\Phi}\Phi )^2
    \right],
  \end{split}
\end{align}
adopted from~\cite{ratti_thermodynamics_2006}. Together with the
parameter $ a(T) = a_0 + a_1 \left( T_0/ T \right) + a_2 \left( T_0/T
\right)^2$ and all parameters therein (see~\cite{rau_chiral_2013} for
values), $U(T,\Phi,\bar{\Phi})$ is constructed such as to reproduce
lattice data for QCD thermodynamics in the pure gauge sector as well
as known features of the deconfinement
transition~\cite{ratti_thermodynamics_2006}. At low temperatures in
the confined phase, the minimum of the potential lies at $\Phi = 0$
and it gradually shifts with higher temperatures to $\Phi \rightarrow
1$ above the critical Polyakov temperature $T_0$.\par
Minimizing $\Omega / V (T,\mu)$ with respect to the fields, yields the
equations of motion of the fields and particle densities. Solving this
set of equations, all thermodynamic variables are derived from the
pressure $p = -\partial \Omega / \partial V$ and the entropy density
$s = \partial p / \partial T $ and the expression for the internal
energy $\epsilon = Ts - pV + \sum_i \mu_i \,\rho_i$, where $i$
includes all particles in the model.\par
With increasing temperatures, the particle density of hadrons
decreases significantly leaving pure quark-gluon matter in the
high-temperature limit. In the chiral model, this shift in the
fundamental degrees of freedom is implemented via an eigenvolume
$V_{\rm ex}^i$ of all hadrons $i$, in analogy
to~\cite{rischke_excluded_1991, mishra_effect_2007,
  steinheimer_effective_2011} and also used in similar hadron
models~\cite{cleymans_excluded_1992, ma_finite_1993,
  yen_excluded_1997, gorenstein_van_1999, bugaev_van_2000}. The
baryons exhibit a volume $V_{\rm ex}^B$ close to the proton charge
volume~\cite{mohr_2010_2011} and the mesons $V_{\rm ex}^M = 1/8\,
V_{\rm ex}^B$. Since quarks are assumed to be point-like $V_{\rm ex}^q
= 0$. This formalism ensures an effective suppression of hadrons at
high $T$ and $\mu$, at the latest when quark abundances rise quickly
at the deconfinement phase transition, and in the high-$T$, high-$\mu$
limit a pure quark-gluon phase is established.  In order not to spoil
the model's thermodynamic consistency, the introduction of an excluded
volume entails the re-definition of the chemical potentials, i.e.\
reducing $\mu^*_i$ by the occupied volume as shown
in~\cite{rau_chiral_2013}. Furthermore, the particle densities as well
as the energy and the entropy have to be corrected by the ratio of the
total volume to the non-occupied sub-volume.

\subsection{Susceptibilities}
\label{sec:susceptibilities}

At any given point in the phase diagram $(T,\mu_B)$, the pressure
$p(T,\mu_B)$ can be determined by Taylor expanding the pressure at $T$
and zero baryonchemical potential $p(T, \mu_B = 0) = -\Omega/V$ with
respect to the ratio $\mu_B/T$
\begin{equation}
  \label{eq:suscep_quarknumber}
  \frac{p(T,\mu_B)}{T^4} = \sum_{n=0}^{\infty} c^B_n(T) \left(
    \frac{\mu_B}{T} \right)^n.
\end{equation}
In the limit of small $\mu_B$, this method yields good numerical
results and is widely used by lattice QCD for the extrapolation of
data at $\mu_B \ne 0$ along lines of constant
$\mu_B/T$-ratio~\cite{karsch_thermodynamics_2003,
  huovinen_fluctuations_2010, schmidt_net-baryon_2010,
  schmidt_conserved_2013}. The Taylor coefficients of the order $n$
are defined as
\begin{equation}
  \label{eq:suscep_cn}
  c^B_n(T) = \left. \frac{1}{n!} \frac{\partial^n \left( p(T,\mu_B) /
        T^4\right)} {\partial \left( \mu_B / T\right)^n
    }\right|_{\mu_B = 0}, 
\end{equation}
which are related to the susceptibilities $\chi^B_n$, analogously to
cumulants in classical statistics, via
\begin{equation}
  \label{eq:relation-coeff-suscep}
  \chi^B_n = n!\, c_n^B. 
\end{equation}
In general, the susceptibilities $\chi^{i,j,k}$ are derived from the
grand canonic partition function $\mathcal{Z} = \exp\left(-\Omega/T
\right)$ using
\begin{equation}
  \label{eq:suscep_partition_fct}
  \chi^{i,j,k}_n (T) = \frac{1}{V\, T^3}
  \frac{\partial^{n_i} \partial^{n_j} \partial^{n_k}} {\partial
    (\mu_i/T)^{n_i}\, \partial (\mu_j/T)^{n_j}\, \partial
    (\mu_k/T)^{n_k}} \ln(\mathcal{Z}).
\end{equation}
The $\chi^{i,j,k}_n$ signal fluctuations of conserved charges
$Q^{i,j,k}$~\cite{karsch_towards_2011}. Considering three-flavor QCD,
the conserved charges $Q$ are baryon number $B$, electric charge $Q$
and strangeness $S$. In the following, only $B$ and $S$ fluctuations
are considered.\par
Following this procedure, the first-order strange quark susceptibility
$\chi_2^S(T)$ is determined by expanding the pressure with respect to
the strange chemical potential $\mu_s$
\begin{equation}
  \label{eq:suscep_strange}
  \chi_2^S(T) = \left. \frac{\partial^2 (p(T,\mu_s))}{\partial
      \mu_s^2} \right|_{\mu_s = 0}.
\end{equation}
This quantity describes fluctuations of the strangeness quantum number
at zero strange quark chemical potential.\par
Since in the chiral hadronic model, $p(T, \mu_B, \mu_s)$ can be
calculated at any given point in the phase diagram, susceptibilities
can directly be determined numerically via
Eqs.~\eqref{eq:suscep_cn},~\eqref{eq:suscep_strange}.

\section{Results}\label{sec:results}

In order to quantify the impact of the quark phase and of repulsive
vector interactions, this study of fluctuations in the transition
region compares susceptibilities with different model
parameterizations differing in the fundamental particle constituents
and the respective couplings strengths. The {\em HRG} scenario
describes the pure HRG in absence of a quark phase neglecting any
excluded volume effects. In this scenario, the hadron resonance gas is
considered to be ideal, i.e.\ all hadronic degrees of freedom do not
couple to the fields. Therefore, the particles' masses are fixed at
their vacuum expectation values. Here, hadrons are considered to be
point-like.\par
The interacting HRG scenario ({\em int.\ HRG}) includes only hadron
degrees of freedom as well. In contrast to the HRG parametrization, in
this scenario hadrons couple to the meson fields as described
above. As a result, their masses and effective chemical potentials are
dynamically generated. Since there is no quark phase, this
parametrization also does not take into account excluded volume
effects.\par
When implementing the PNJL-like quark phase, the {\em int.\ HRG+q}
parametrization denotes the best practice scenario including hadrons
and quarks fully coupled to the fields and hadrons exhibiting a finite
eigenvolume.\par
The study of the strange susceptibility additionally makes use of the
{\em HRG+q} parametrization. This scenario contains hadrons and quarks
which are considered to be ideal, i.e.\ all couplings to the meson
fields vanish. However, quarks still couple to $\Phi$ and excluded
volume effects apply.

\subsection{Non-Strange Susceptibilities}
\label{sec:non-strange-susceptibilities}

\begin{figure}[tb]
  \centering
  \includegraphics[width=1.\columnwidth]{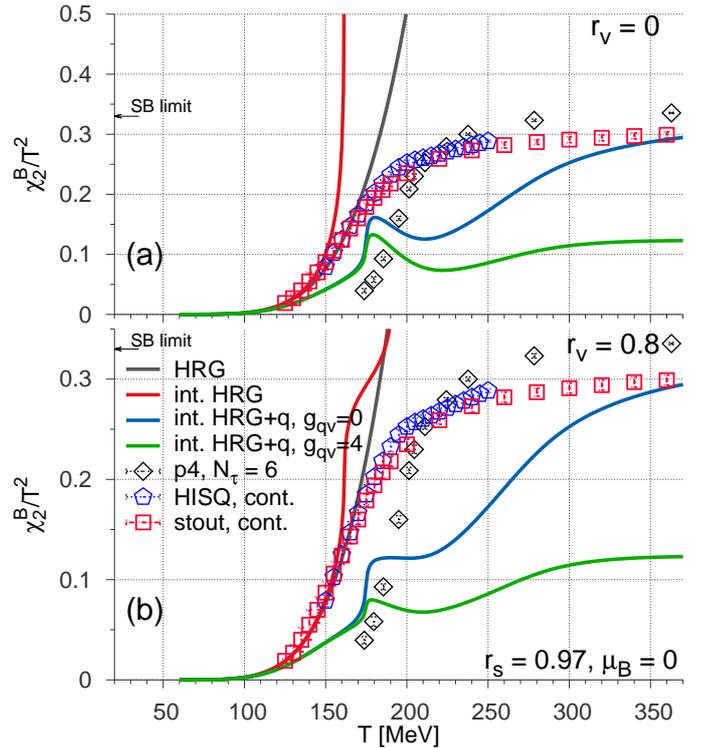}
  \caption{Second-order baryon number susceptibilities $\chi^B_2/T^2$
    at $\mu_B = 0$ as function of $T$. Depicted are results for
    resonance vector couplings $r_v = 0$ (a) and $r_v = 0.8$ (b) for
    the non-interacting pure HRG (gray line) and the fully interacting
    HRG (red line). Furthermore, results of the fully interacting
    model including quarks are shown for quark vector couplings
    $g_{\rm q v} = 0$ (blue line) and $g_{\rm q v} = 4.0$ (green
    line). The quark scalar coupling is fixed at $g_{\rm q s} =
    3.5$. Strong vector coupling suppress fluctuations
    significantly. Considering a finite quark vector coupling,
    fluctuations get unreasonably restrained above $T_c$. Lattice data
    is taken from~\cite{bazavov_fluctuations_2012}
    (HISQ),~\cite{Cheng:2008zh} (p4), and~\cite{Borsanyi:2011sw}
    (stout).}
  \label{fig:susceptibilities-c2-mu-zero}
\end{figure}

Figure~\ref{fig:susceptibilities-c2-mu-zero} shows the second-order
baryon number susceptibilities $\chi^B_2/T^2$ at $\mu_B = 0$ as
functions of $T$ contrasted to lattice QCD data using different
actions~\cite{bazavov_fluctuations_2012, Cheng:2008zh,
  Borsanyi:2011sw}. Panel (a) depicts the susceptibilities for a
vanishing resonance vector coupling $r_v = 0$ and (b) for $r_v =
0.8$.\par
As a reference, the gray line shows the non-interacting HRG without
quarks. Since this scenario lacks a phase transition, there is no
shift in the underlying degrees of freedom. Hence, there is no sudden
change in $\chi^B_2$ but rather it rises monotonously with increasing
temperature due to the gradual population of heavy-mass resonance
states. This monotonic behavior is found for both resonance vector
couplings regarded here. In this scenario, the absence of suppressive
vector field interactions causes an overestimation of the number of
degrees of freedom entailing large $\chi^B_2/T^2$-values at high
$T$.\par
This overestimation is even enhanced for the fully interacting HRG
(red line). Considering full scalar field interactions and vanishing
resonance vector couplings (a), starting from $T \approx 150$~MeV the
slope of $\chi^B_2$ is much steeper than for the non-interacting
HRG. This rapid rise of susceptibilities in the interacting scenario
signals the sudden decline of effective baryon masses at higher $T$ at
which $m^*$ becomes small enough for massive resonance states to be
abundantly populated. In a small range around $T_c \approx
165$~MeV~\cite{rau_baryon_2012} the mass of the $\Delta$-resonances
falls down to $m^*_{\Delta} \approx 0.4\, m^*_{\Delta}(T_0)$ and in
the absence of a quark phase $\Delta$-resonances become most abundant
above $T_c$~\cite{rau_chiral_2013}. The drop of $m^*$ causes a sudden
increase of degrees of freedom which is reflected in the steep incline
of $\chi^B_2$ at $T_c$.\par
In contrast, when choosing larger and more reasonable resonance vector
couplings $r_v = 0.8$~\cite{rau_baryon_2012}, hadron degrees of
freedom are notably suppressed above $T_c$ as depicted in
Fig.~\ref{fig:susceptibilities-c2-mu-zero} (b). With $r_v = 0.8$ the
drop of $m^*$ is still existent and $\chi^B_2$ of the int.\ HRG shows
a sudden increase at $T_c$. However, at slightly higher temperatures
$\chi^B_2/T^2 (T)$ flattens and rises again at $T \approx 180$~MeV,
where the $\zeta$-field drops and strange baryon masses become
light. In contrast, the light quark susceptibility $\chi^{u,d}_2$,
which has no strange contribution, saturates at $\chi^{u,d}_2/T^2
\approx 1$ slightly above $T_c$. This finding underlines the major
impact of hadron resonances and their couplings on the phase structure
and the overall behavior of the system at $T_c$ found
in~\cite{rau_baryon_2012, rau_chiral_2013}. The key implication of
these rather large baryon resonance vector couplings lies in the
disappearance of a first order phase transition and a critical end
point. With $r_v \to 1$ only a smooth cross-over exists in the whole
phase diagram~\cite{rau_baryon_2012}.\par
Comparing the purely hadronic results to lattice QCD (data points
corresponding to different lattice actions), the slope of
$\chi^B_2/T^2(T)$ from recent continuum extrapolated lattice
QCD~\cite{bazavov_fluctuations_2012, Borsanyi:2011sw} is in line with
HRG results of the model up to $T \approx 150$--170~MeV depending on
the couplings. Older lattice results with the p4
action~\cite{Cheng:2008zh} show a slightly higher $T_c$ and seem to be
too small at $T < T_c$ compared to HRG results. This discrepancy may
be caused by potential cut-off effects in lattice QCD on small lattice
sites ($N_\tau \le 8$).\par
Similar observations for the suppression of fluctuations with stronger
vector interactions are made for the fully interacting model including
quarks. In the presence of a quark phase, quark vector couplings have
a large effect on the $\chi^B_2$-slope. In both panels,
Fig.~\ref{fig:susceptibilities-c2-mu-zero} shows results of the chiral
model including quarks with vanishing quark vector couplings $g_{q
  \omega} = 0$ (blue line) and with finite couplings $g_{q \omega} =
4.0$ (green line), with the notation $g_{q \omega}$ and $g_{q v}$ used
synonymously. In both HRG+q scenarios the second-order
susceptibilities exhibit a peak at $T_c \approx 175$~MeV when the
shift in the degrees of freedom is the fastest. For higher $T$,
susceptibilities for $g_{q \omega} = 0$ rise to the Stefan-Boltzmann
(SB) limit while they saturate at much lower values in the case of
$g_{q \omega} = 4.0$.  At high temperatures, excluded volume effects
cause an effective suppression of hadrons and result in a plateau-like
slope or even a small decline of $\chi^B_2$ up to $T \approx
220$~MeV. In the presence of quarks, the effect of the resonance
vector couplings on the susceptibilities lessens and when changing
$r_v = 0$ (a) to $r_v = 0.8$ (b), the absolute height is only reduced
by a small amount for both $g_{q \omega}$-values.\par
In contrast to this rather small impact of the hadron vector couplings
in the presence of a quark phase, the quark vector couplings $g_{q
  \omega}$ have a strong quenching effect on fluctuations. When $g_{q
  \omega}$ changes from zero to $g_{q \omega} = g_{N \omega}/3 = 4.0$,
the height of the peak in $\chi^B_2$ decreases significantly and,
likewise, the deviation from lattice data increases for $g_{q \omega}
= 4.0$ at high $T$. In~\cite{lastowiecki_neutron_2012,
  blaschke_nonlocal_2013} it is argued, that a large quark vector
coupling is needed to properly describe heavy-mass neutron stars
within the framework of a PNJL equation of state (EoS). This reasoning
bases on the stiffening of the EoS with larger quark vector
couplings~\cite{rau_chiral_2013}. Due to this substantial stiffening,
the mass-radii relation for neutron stars is shifted towards higher
masses, matching a constraint put up by the recent observation of
massive two-solar-mass hybrid
stars~\cite{demorest_two-solar-mass_2010}. In stark contrast to this
constraint on the EoS from neutron star properties, in the context of
conserved charge fluctuations, non-vanishing quark vector couplings
must be ruled out due to the strong suppression of susceptibilities
above $T_c$. Above $T_c$, $\chi^B_2$ values from lattice QCD are
achieved only in the full model without quark vector interactions.
The suppression of fluctuations in the transition region is even more
pronounced for higher-order susceptibilities.\par
The comparison of the Polyakov loop from the chiral model to lattice
QCD as a function of the temperature~\cite{rau_chiral_2013} indicates
a rather slow and smooth shift in the degrees of freedom over a large
temperature-range in lattice QCD rather than a more sudden switching
to quarks and gluons at $T_c$ in PNJL models and in the chiral model.
This discrepancy is underlined by the deviation of $\chi^B_2$ in the
chiral model from lattice QCD. While the fully interacting HRG in the
chiral model shows a sharp incline of $\chi^B_2$ at $T_c$, in lattice
QCD the absence of a sharp rise in the susceptibilities and an
increasing $\chi^B_2$ up to rather large temperatures again hint at a
more gradual shift from hadrons to quarks.  This immanent discrepancy
between lattice QCD and PNJL models in the transition to the quark
sector is reflected in conserved charge fluctuations close to the
critical temperature. In~\cite{ratti_are_2012} the differences between
PNJL and lattice QCD results have been attributed to the presence of
bound states in the QGP even well above $T_c$. However, in the chiral
model fluctuations above $T_c$ are suppressed by excluded volume
effects of baryons which are still present in this region.\par
The same conclusions as for $\chi^B_2$ also apply for higher-order
susceptibilities and the impact of vector couplings on the occurrence
of fluctuations close to $T_c$.
\begin{figure}[tb]
  \centering
  \includegraphics[width=1.\columnwidth]{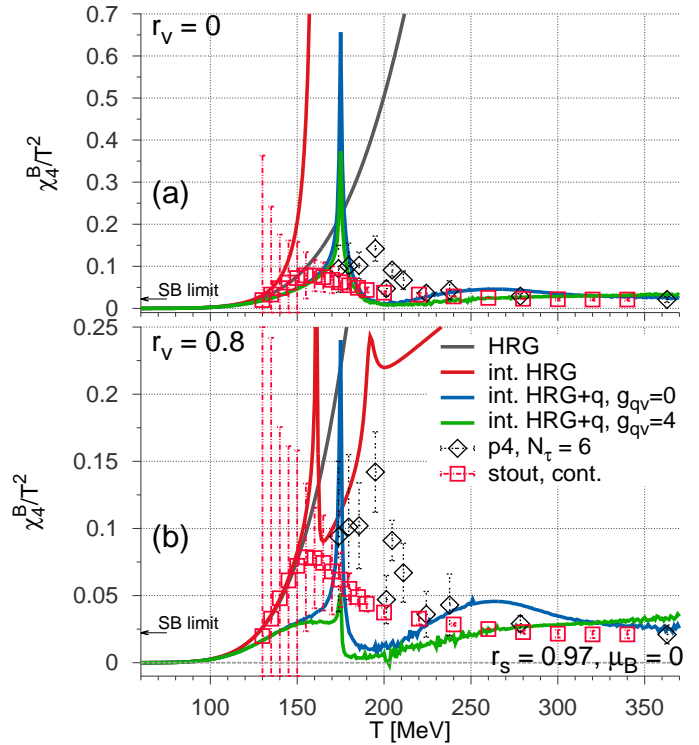}
  \caption{Fourth-order baryon number susceptibilities $\chi^B_4/T^2$
    at $\mu_B = 0$ as functions of $T$. The figure depicts results for
    two values of $r_v$ (a) and (b) and for the same model scenarios
    as in Fig.~\ref{fig:susceptibilities-c2-mu-zero}. Fluctuations at
    the phase transition decrease significantly when choosing larger
    vector couplings of the particles. For non-zero quark vector
    couplings $g_{qv}$ (green line in (b)), $\chi^B_4/T^2$ exhibits
    only a small maximum at $T_c$. Lattice data
    from~\cite{Cheng:2008zh} (p4) and~\cite{Borsanyi:2011sw,
      Borsanyi:2013hza} (stout).}
  \label{fig:susceptibilities-c4-mu-zero}
\end{figure}
Figure~\ref{fig:susceptibilities-c4-mu-zero} shows the fourth-order
susceptibilities $\chi^B_4/T^2$ for the same model scenarios as
above. As seen for $\chi^B_2$, $\chi^B_4$ of the non-interacting HRG
without quarks rises smoothly up to very high values. For the
interacting HRG and $r_v = 0.8$, $\chi^B_4(T)$ exhibits a sharp peak
at $T_c$ and overestimates the susceptibilities at higher $T$. For
$r_v = 0.8$, the contribution of strange baryons already seen in
$\chi^B_2$ leads to an additional peak at $T \approx
190$~MeV. Including the quark phase with $g_{q v} = 0$ (blue line) and
$r_v = 0.8$, this scenario (int.\ HRG+q) yields a sharp and narrow
peak around $T_c$.  In contrast, the more recent and continuum
extrapolated stout data, which are derived from~\cite{Borsanyi:2011sw}
and~\cite{Borsanyi:2013hza}, indicate a much broader range of
fluctuations which agrees with HRG results at low temperatures and
reaches the quark limit at higher $T$.  In the high-temperature limit,
lattice QCD and model results including quarks converge to the SB
limit.\par
As for $\chi^B_2$, a larger $g_{q v}$ strongly suppresses
fluctuations. When assuming $g_{q \omega} = 4.0$, the resulting
$\chi^B_4(T)$ shows only a minor peak at $T_c$. This confirms the
observation, that $g_{q \omega}$ has to vanish in order to reproduce
fluctuations as determined by lattice QCD. In contrast to $\chi^B_2$,
$\chi^B_4$-values from both int.\ HRG+q scenarios yield similar values
in the high-$T$ limit.\par
In heavy-ion collisions, susceptibilities are subject to additional
random fluctuations due to changing volumes of the colliding systems
caused by randomly different collision geometries in each collision
process. To circumvent this constraining effect and to consistently
remove the impact of ever varying volumes, susceptibility ratios are
studied~\cite{koch_hadronic_2008}. In~\cite{ejiri_hadronic_2005} it is
shown that the ratio of fourth to second-order susceptibilities
$\chi^B_4/\chi^B_4$ is sensitive to the shift in the system's
underlying degrees of freedom. This ratio can provide information
about the constituents of a thermal medium that carries net quark
number in both the HRG as well as in the quark phase.\par
\begin{figure}[tb]
  \centering
  \includegraphics[width=1.\columnwidth]{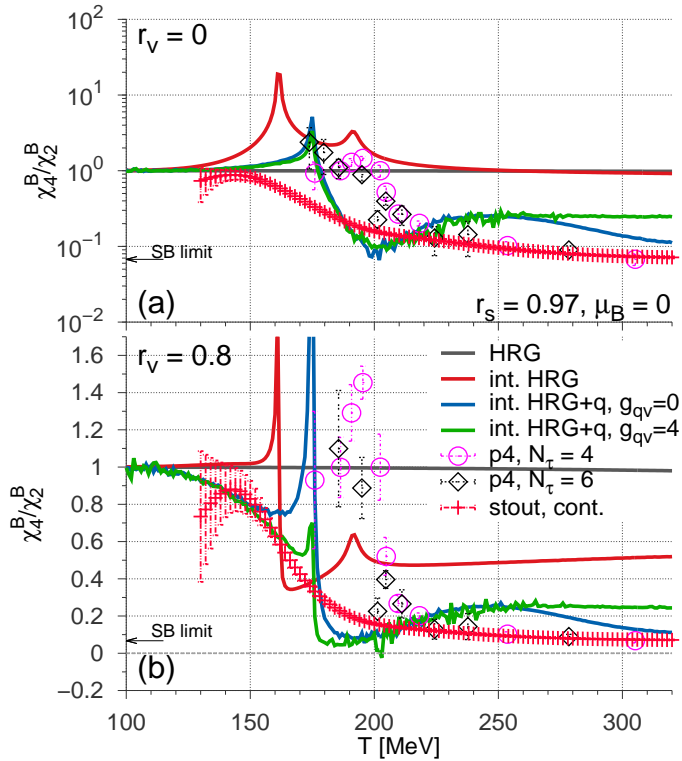}
  \caption{Ratio of the fourth to the second-order baryon number
    susceptibilities $\chi^B_4/\chi^B_2$ at $\mu_B = 0$ as functions
    of $T$. As before, the figure illustrates results for two
    resonance vector couplings $r_v = 0$ (a) and $r_v = 0.8$ (b) and
    for model scenarios as in
    Fig.~\ref{fig:susceptibilities-c2-mu-zero}.  Lattice data taken
    from~\cite{Cheng:2008zh} (p4) and~\cite{Borsanyi:2013hza}
    (stout).}
  \label{fig:ratios-mu-zero}
\end{figure}
Figure~\ref{fig:ratios-mu-zero} shows the ratio $\chi^B_4/\chi^B_2$
from the model at $\mu = 0$ as a function of $T$. As seen for
$\chi^B_2$ and $\chi^B_4$, the ratio signals a rapid shift in the
degrees of freedom at $T_c$ by a narrow peak for all scenarios
considering full mean field interactions. Again, the effective hadron
suppression via excluded volume effects as well as via resonance
vector couplings for all particles have major impact on the
fluctuation strength and hence, on the height of the peak at
$T_c$. Therefore, the HRG+q scenario assuming $g_{q \omega} = 4.0$
(green line) clearly underestimates $\chi^B_4/\chi^B_2$ at $T_c$.\par
None of the scenarios presented here fully reproduces
$\chi^B_4/\chi^B_2$-results from lattice QCD which again show a rather
narrow peak for older p4 and a much broader peak for recent stout
results. At high temperatures $T > T_c$, effective model results with
$g_{q \omega} = 0$ and lattice QCD converge again in the quark limit.

\subsection{Strange Susceptibility}
\label{sec:strange-susceptibility}

Below the deconfinement transition in the HRG, strangeness is carried
by strange hadrons, mainly strange mesons (kaons). Compared to the
temperature scale in this region, strange hadrons have rather large
masses. Therefore, the production of strange hadrons is largely
suppressed and their multiplicities are small at low
$T$~\cite{rau_baryon_2012, rau_chiral_2013}. Contrastingly, in the
high-$T$ limit, i.e.\ in the pure quark-gluon phase, low-mass strange
quarks contribute exclusively to the total strangeness in the
system. Due to the much smaller quark masses, fluctuations in the
strangeness number should increase rapidly at the transition from
hadrons to quarks. In the pure quark-gluon phase fluctuations of the
strange quark number should reach a maximum. Due to this direct
connection of strange quark number fluctuations with the underlying
degrees of freedom, the strange quark susceptibility $\chi_2^S$
signaling strangeness fluctuations might serve as an indicator for the
deconfinement transition (see e.g.~\cite{Cheng:2008zh,
  aoki_qcd_2009, borsanyi_is_2010, Borsanyi:2011sw,
  bazavov_fluctuations_2012, schmidt_conserved_2013} for $\chi^s_2$ from
lattice QCD).\par
\begin{figure}[tb]
  \centering
  \includegraphics[width=1.\columnwidth]{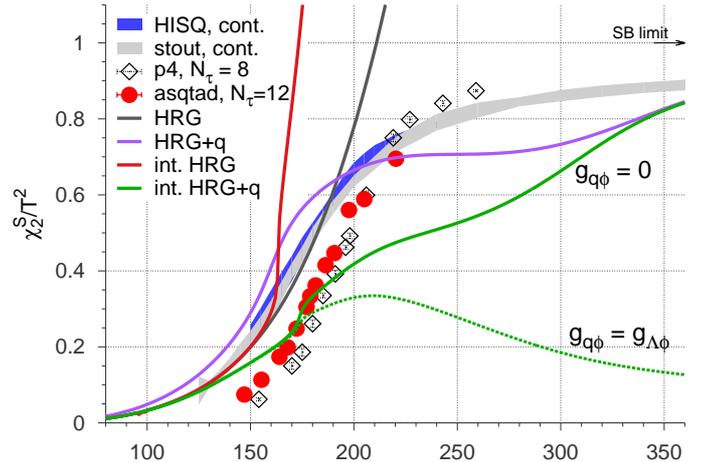}
  \caption{Strange quark number susceptibilities $\chi^s_2/T^2$ at
    $\mu_B = 0$ as functions of $T$ for the different model
    scenarios. The model results are contrasted to lattice data
    from~\cite{Borsanyi:2011sw, bazavov_deconfinement_2010,
      bazavov_chiral_2011-1, borsanyi_is_2010, borsanyi_qcd_2010,
      aoki_qcd_2009, endrodi_qcd_2011, schmidt_conserved_2013,
      bazavov_fluctuations_2012}. Here, the resonance vector coupling
    is set to $r_v = 0$. When the vector coupling of the strange quark
    increases from $g_{q\phi} = 0$ (solid green line) to $g_{q\phi} =
    g_{q\Lambda} = -7.4$ (dashed green line), strange quark number
    fluctuations are notably suppressed at high $T$.}
  \label{fig:strange-quark-suscep}
\end{figure}
Figure~\ref{fig:strange-quark-suscep} depicts the strange
susceptibility $\chi_2^S$ divided by $T^2$ as a function of
$T$. Although restricting the net strangeness in the total system to
$f_s = 0$, this quantity reflects the fundamental difference in
underlying degrees of freedom between the different model scenarios
considered here. In the non-interacting HRG, more heavy-mass strange
hadrons are produced with increasing $T$. Since there is no abrupt
shift in the degrees of freedom and the hadron masses do not change
due to the absence of mean field interactions, $\chi_2^S$ rises
continuously. The $\chi_2^S(T)$ reproduces lattice results up to
$T_c$. The behavior of $\chi_2^S$ changes when $m^*$ drops at $T_c$ due
to non-vanishing scalar field couplings. In this case (red line), with
increasing $T$ baryons loose a large amount of their mass and
significantly more strange hadrons are produced with higher $T$.
Hence, in this case $\chi_2^S$ exhibits a steep and sudden rise at the
critical temperature.\par
Next, this study turns to the additional quark phase and its effect on
$\chi_2^S$ at the phase transition. With the HRG+q scenario neglecting
all mean field interactions (purple line), the shift in the degrees of
freedom is reflected by an increase of $\chi_2^S$ with a curvature
qualitatively comparable to lattice results. When quarks dominate the
system at temperatures above $T \approx 1.5 \, T_c$, $\chi_2^S$
flattens. Compared to other model scenarios, in this non-interacting
HRG+q scenario $\chi_2^S$ is significantly larger at $T < T_c$. This
effect can be attributed to the lack of repulsive quark interactions
causing quarks to be present even at very low $T$ and, hence,
increasing strange quark fluctuations in this region. The appearance
of quarks below $T_c$ notwithstanding, it is shown
in~\cite{rau_chiral_2013} that even a small quark vector couplings can
prevent free quarks from populating the ground state and reasonable
ground state properties can be reproduced within the effective
model.\par
Considering the int.\ HRG+q scenario with full mean field interactions
(green lines) the slope of $\chi_2^S(T)$ changes.
Figure~\ref{fig:strange-quark-suscep} shows results for vanishing
vector couplings of the strange quarks $g_{q \phi} = 0$ (solid green
line) and for a finite value $g_{q \phi} = g_{\Lambda \phi} = -7.4$
(dashed green line). For both couplings, $\chi_2^S(T)$ is much flatter
as in the HRG+q scenario without interactions due to a prolonged shift
from hadrons to quarks. When neglecting strange quark vector
interactions $g_{q \phi} = 0$, the value of $\chi_2^S$ reaches the ideal
quark gas limit (purple line) at high temperatures $T \approx 2\,
T_c$. Contrastingly, when strange quarks are suppressed by vector
field interactions, $\chi_2^S$ reaches its maximum at $T_c$, where
strange hadrons and quarks coexist and decreases again due to the
vector suppression of strange degrees of freedom with higher
temperatures.\par
This finding underlines the observations made for non-strange
susceptibilities: Even small vector field interactions cause a
significant suppression of fluctuations at the phase
transition. Comparing susceptibilities from the model to lattice QCD
results, it becomes apparent, that strange quarks can not exhibit a
considerable coupling to the respective vector field.\par
When regarding the strange susceptibility as an observable for
deconfinement, one should keep in mind that not only a shift from
hadrons to quarks but also the drop in the effective strange-hadron
masses leads to a sudden increase of $\chi_2^S$ at the critical
temperature. For this reason, if the chiral and the deconfinement
phase transition do not happen at the same temperature, fundamentally
different and more complicated $\chi_2^S(T)$-curves with considerable
contributions both from strange hadrons and from strange quarks near
the the phase transition might be possible.

\section{Summary and Conclusions}
\label{sec:sum-and-conclusio}

This work presents non-strange and strange susceptibilities at the
phase transition ($T_c \approx 165$~MeV) at $\mu_B = 0$ obtained from
a chiral model including a PNJL-like quark phase. The model includes
all known hadrons and quarks and by this features a chiral transition
as well as deconfinement.\par
Comparing model results to lattice QCD, it shows that heavy-mass
baryon resonances have a large effect on the hadronic sector of quark
number fluctuations at $T_c$ and the steep increase of
susceptibilities in this region results from the occurrence of
multiple baryon resonance states. However, rather large repulsive
vector field interactions for baryon resonances must be taken into
account in order to restrain light quark number susceptibilities to
values found in lattice QCD. In $\sigma$-$\omega$ models including a
large spectrum of heavy-mass resonances strong vector field
interactions necessarily lead to moving the critical end point to
large $\mu_B$. In the chiral model used here, vector field
interactions of the size determined in this study cause the first
order phase transition and a critical end point to
vanish~\cite{rau_baryon_2012}.\par
In the quark sector, model results show that particles are almost
acting like an ideal gas. Particularly, this implies that repulsive
interactions of quarks with strange and non-strange vector meson
condensates must vanish in order not to annihilate fluctuations above
$T_c$. The strange quark number susceptibility reflects not only the
shift from hadrons to quarks but also signals a large contribution
from strange baryons, which loose most of their effective masses at
the chiral transition. Therefore, the strange susceptibility should
not be regarded as a clear indicator for the shift in the degrees of
freedom. This finding encourages further model studies to explore the
contribution of hadrons and quarks at the transition using the
baryon-strangeness correlator to show how fast the ideal gas limit is
achieved.

\section{Acknowledgements}
\label{sec:ack}

Work was supported by BMBF, GSI, and by the Hessian excellence
initiative LOEWE through the Helmholtz International Center for FAIR
(HIC for FAIR), and the Helmholtz Graduate School for Hadron and Ion
Research (HGS-HIRe). Computational resources were provided by the
Center for Scientific Computing (CSC) of the Goethe University
Frankfurt. J.~S.\ acknowledges a Feodor Lynen fellowship of the
Alexander von Humboldt foundation. The authors thank the
Wuppertal-Budapest collaboration for providing recent data on
$c_4/c_2$ and F.\ Karsch and the HotQCD collaboration for fruitful
discussion.

\bibliographystyle{apsrev}
\bibliography{new}

\end{document}